\documentclass[pre,preprint,aps,eqsecnum]{revtex4}
\usepackage{epsfig}

\begin{document}

\title{Deformation and Failure of Amorphous Solidlike Materials}

\author{Michael Falk}
\affiliation{Department of Materials Science and Engineering, Department of Mechanical Engineering and Department of Physics and Astronomy, Johns Hopkins University, Baltimore, MD 21218}

\author{J.S. Langer}
\affiliation{Department of Physics, University of California, Santa Barbara, CA  93106-9530}

\date{\today}

\begin{abstract}
Since the 1970's, theories of deformation and failure of amorphous, solidlike materials have started with models in which stress-driven, molecular rearrangements occur at localized flow defects via ``shear transformations.''  This picture is the basis for the modern theory of ``shear transformation zones'' (STZ's), which is the focus of this review. We begin by describing the structure of the theory in general terms and by showing several applications, specifically: interpretation of stress-strain measurements for a bulk metallic glass, analysis of numerical simulations of shear banding, and the use of the STZ  equations of motion in free-boundary calculations.  In the second half of this article, we focus for simplicity on what we call an ``athermal'' model of amorphous plasticity, and use that model to illustrate how the STZ theory emerges within a systematic formulation of nonequilibrium thermodynamics.  

\end{abstract}
\maketitle

\section{Overview}
\label{overview}

Deformation and failure of amorphous, solidlike materials is a large topic with a long history. This class of materials includes structural and metallic glasses, glassy polymers, dense colloidal suspensions, many kinds of granular materials, and a huge range of biological substances.   Despite the challenge presented by their inherent disorder, these materials have received special attention because of the broad range of applications in which it is essential to predict their failure modes.  The ways in which amorphous solids behave when driven by external forces have been studied throughout the latter half of the 20'th century by disparate groups of physicists and materials scientists. The field has advanced significantly in recent years, largely because of insights gained from numerical simulations \cite{FALK-MALONEY-10}, and partly because of the introduction of a few new theoretical concepts; but even the basic form of an acceptable, first-principles theory has remained a matter of debate.

Among the computation-inspired advances is the ``shear-transformation-zone'' (STZ) theory of amorphous plasticity, which we proposed in 1998 \cite{FL98}, and which has gone through a series of modifications and extensions since then.  (For example, see \cite{LP03,FLP-04,JSL04,PECHENIK05,BLP07I,BLP07II,JSL08}.)  The STZ theory is the principal topic of this review.  We emphasize two main points: first, that the STZ theory starts with specific assumptions about the nature of molecular rearrangements to arrive at testable predictions of experimental data; and, second, that the theoretical hypotheses are strengthened by being developed in the context of fundamental principles of nonequilibrium thermodynamics. 

From its inception \cite{FL98}, the STZ theory was intended to be an extension of the flow-defect theories of Turnbull, Cohen, Spaepen, Argon and others \cite{TURNBULL-COHEN70,SPAEPEN77,ARGON79,SPAEPEN81,ARGON83} in which localized clusters of molecules undergo irreversible rearrangements in response to applied shear stresses. In a very rough sense, the flow defects in amorphous materials play the role of dislocations in crystals by being the agents of plastic deformation.  All of these theories, including STZ, start by assuming that the material of interest is solidlike -- that it has a shear modulus -- and that the flow defects or dislocations allow it to behave in some respects like a liquid.  Thus, the flow-defect theories are qualitatively different from fluid-based theories such as mode-coupling \cite{GOTZE91,GOTZE92,MC-Cates-09}, which starts from a liquidlike, many-body description and predicts the onset of solidlike behavior at high densities and low temperatures.   

The most important way in which the STZ theory departs from its predecessors is by recognizing that the flow defects must possess internal degrees of freedom. The STZ's not only appear and disappear during configurational fluctuations; they also transform from one orientation to another, and the net rate at which these transformations occur determines the rate of irreversible shear deformation.  As we show later in this review, the equations of motion for this orientational degree of freedom predict that the system undergoes an exchange of dynamic stability from jammed to flowing states at a yield stress or, more accurately, at a minimum flow stress.  This dynamic interpretation of yielding is not exactly the same as the conventional picture of stress-driven barrier crossing; it has as much to do with entropy generation as it does with deterministic dynamics.  The orientational degree of freedom also gives the system a directional memory; for example, it produces Bauschinger effects. \cite{DIETER-86} 

In recent years, the STZ theory has acquired a thermodynamic complexion with the introduction of an effective disorder temperature.  Like the flow-defect picture, the effective temperature idea has historical roots in Cohen and Turnbull's theories of glasses \cite{COHEN-TURNBULL59}.  Other investigators, notably  Spaepen \cite{SPAEPEN77}, described the intrinsically disordered state of noncrystalline materials by a free volume $v_f$.  Those authors perceptively recognized that the relevant definition of $v_f$ is not as an extensive excess volume measured from some densely packed state, but as an intensive quantity -- the inverse of the derivative of a configuratio
nal entropy (i.e. a dimensionless entropy associated with the mechanically stable molecular positions, without kinetic or vibrational contributions) with respect to the volume.  Thus they proposed that the density of flow defects might be proportional to a Boltzmann-like factor, $\exp\,(- {\rm constant}/v_f)$, and not just to $v_f$ itself. This thermodynamic analogy was further developed for strictly athermal materials such as powders by Edwards and coworkers \cite{EDWARDS-89,MEHTA-EDWARDS-90}.  Lemaitre \cite{LEMAITRE-02,LEMAITRE-02a} has used the free volume in a way that is even more closely closely related to the effective-temperature analysis described here.

Whenever the mechanical behavior of the system is determined by interactions between its elementary ingredients, it becomes important to couch the thermodynamic analysis in terms of energy rather than volume. Then the analog of the free volume $v_f$ is an effective disorder temperature $T_{e\!f\!f}$, defined here to be the derivative of the configurational energy with respect to the configurational entropy.  $T_{e\!f\!f}$ equilibrates with the ambient temperature $T$ at high $T$, but may fall out of equilibrium at low $T$ when molecular disorder is generated by  mechanical deformation, or when glass-forming liquids are quenched through a glass transition.  In the latter case, $T_{e\!f\!f}$ is the same as the fictive temperature \cite{TOOL-46,ANGELL-00}.  In analogy to the free-volume formula, the steady-state STZ density is proportional to $\exp\,(-T_Z/T_{e\!f\!f})$, where $e_Z = k_B\,T_Z$ is a characteristic STZ formation energy.

Throughout this review, we consider only materials in which localized molecular rearrangements occur in the presence of an ambient, high-frequency noise generated by either thermal or mechanical fluctuations.  Thus, we do not consider strictly athermal situations that may occur in granular or colloidal materials very near the jamming transition.  In particular, the physics we discuss is quite different from that observed in athermal quasi-static (AQS) numerical simulations \cite{MALONEY-LEMAITRE-04a,MALONEY-LEMAITRE-04b,LEMAITRE-CAROLI-07,LEMAITRE-CAROLI-09,MALONEY-ROBBINS-09}, in which the system is relaxed to its nearest energy minimum after each of a sequence of small strain steps.  Both AQS simulations, and fully dynamic simulations in the AQS limit of vanishingly small temperature and strain rate, characteristically exhibit system-spanning, crack-like events, and size-dependent spectra of stress fluctuations.

Despite the obvious discrepancy between AQS and normal behavior, Lerner and Procaccia \cite{LERNER-PROCACCIA-09} have asserted  that the system-spanning events observed at or near the AQS limit cast serious doubt on all plasticity theories that are based on the assumption of localized events.  We always have held that such behavior may be typical of slowly driven, nearly jammed, granular materials; but conventional, amorphous molecular materials exhibit no such size dependent behavior.  Since the models being simulated have normal, finite-ranged, molecular interactions, it is clear that there must be a crossover from AQS to normal behavior as a function of increasing temperature and strain rate.  The system-spanning events must be suppressed by thermal and mechanical noise; and constitutive relations and fluctuation spectra must behave properly in the thermodynamic limit of infinite volume. More recently, Procaccia and colleagues \cite{HENTSCHELetal-10} have demonstrated that this is indeed the case; in fact, they find that the expected crossover to localized events and normal plasticity occurs at exceedingly low temperatures. This observation makes it clear that it is inappropriate to overgeneralize the AQS results as in \cite{LERNER-PROCACCIA-09}, and reaffirms that the existence of thermal or mechanical noise places amorphous molecular systems -- to a very good first approximation -- squarely in the regime where local events control dynamics, consistent with the STZ theory.
 
A description of amorphous plasticity that we believe must be closely related to the STZ theory is known as ``Soft Glassy Rheology'' or ``SGR'' \cite{SGR-97,SOLLICH-98}.  This theory describes a broad range of glassy behaviors in terms of distributions of localized,  noise activated, displacement processes.  Like the STZ theory, SGR exhibits  transitions between different kinds of jammed and flowing states and can account, at least qualitatively, for a variety of phenomena observed in soft materials.  However, SGR begins by assuming a fixed, broad distribution of activation energies and the existence of a noise temperature that controls the activation rates. The meaning and dynamics of the SGR noise temperature remain unclear at present; we do not even know whether it might be same as the STZ effective temperature.  In contrast, the STZ theory is based on a specific model of molecular rearrangements that have been observed directly in numerical simulations and analog experiments.  Both the effective temperature and the dissipative processes that generate it are defined in terms of those molecular degrees of freedom. 

This review consists of two main parts, presented in Secs.\ref{part1} and \ref{part2}.  In Sec.\ref{part1}, we exhibit one limiting form of the STZ equations of motion and very briefly describe several applications, specifically: interpretation of stress-strain measurements for a bulk metallic glass, analysis of numerical simulations of shear banding, and the use of the STZ  equations in free-boundary calculations.  In Sec.\ref{part2}, we focus for simplicity on what we call an ``athermal'' model of amorphous plasticity, and use that model to illustrate the thermodynamic basis of the STZ theory.  The thermodynamic arguments are taken primarily from \cite{BLI-09,BLII-09,BLIII-09}; but some features of this analysis have not appeared elsewhere.  The derivations in Sec.\ref{part2} provide a first-principles rationale for the formulas used in Sec.\ref{part1}.

\section{Summary and Selected Applications of the STZ Theory}
\label{part1}

\subsection{Elasto-Plasticity}
\label{elastoplasticity}

Our first step in summarizing the STZ theory is to place it within a general set of Eulerian equations of motion for elasto-plastic deformation in a solidlike material.  Consider a $d$ dimensional system; let $i,j...$ be spatial indices; and use summation convention. Then write the stress tensor $\sigma_{ij}$ in the form:
\begin{equation}
\label{sp}
\sigma_{ij}=-p\,\delta_{ij}+s_{ij},~~~p=-\frac{1}{d}\,\sigma_{kk},
\end{equation}
where $p$ is the pressure and $s_{ij}$ is the traceless, symmetric, deviatoric stress.  In analogy to fluid dynamics, let $v_i(x,t)$ denote the material velocity at the physical position $x=\{x_i\}$ and time $t$.  The acceleration and continuity equations are: 
\begin{equation}
\label{acceleration}
\rho_0\,\frac{dv_i}{ dt}={\partial\sigma_{ij}\over\partial x_j}=-{\partial p\over\partial x_i}+{\partial s_{ij}\over \partial x_j};~~~~
\frac{d\rho_0}{dt}= - {\partial\over \partial x_i}\,(\rho_0\, v_i).
\end{equation}
Here, $\rho_0$ is the density, and $d/dt$ denotes the material time derivative
acting on a scalar or a displacement or velocity field:
\begin{equation}
\frac{d}{dt}\equiv {\partial\over \partial t}+ v_k\,{\partial\over \partial x_k}.
\end{equation}
For most purposes, we assume that $\rho_0$ remains approximately constant and thus do not need the second, continuity equation in Eq.(\ref{acceleration}). 

The problem of disentangling elastic and plastic deformations has long been one of the more serious challenges in solid mechanics.  This problem is not our main topic.  (See \cite{ANAND-GURTIN} for a recent review.)  Rather than address it, we assume -- as a simple working approximation -- that the total rate-of-deformation tensor $D^{tot}_{ij}$ can be written as a linear superposition of elastic and plastic contributions:
\begin{equation}
\label{Dtensor}
D^{tot}_{ij}
\equiv \frac{1}{2}\,
\left({\partial v_i\over \partial x_j}+{\partial v_j\over \partial x_i}\right)
=\frac{\mathcal{D}}{\mathcal{D}t}\,
\left(-\frac{p}{2K}\,
\delta_{ij}+ \frac{1}{2\mu}\,s_{ij}\right)
+ D^{pl}_{ij},
\end{equation}
where $\mu$ is the shear modulus, $K$ is the inverse compressibility, and $D^{pl}_{ij}$ is the plastic rate of deformation. The symbol $\mathcal{D}/\mathcal{D}t$ denotes the material time derivative acting on a tensor, say $A_{ij}$.  For present purposes, we can write:
\begin{equation}
\label{timederivative}
\frac{\mathcal{D}A_{ij}}{\mathcal{D}t}
\equiv {\partial A_{ij}\over \partial t}+v_k\,{\partial A_{ij}\over \partial x_k}+ A_{ik}\,
\omega_{kj} -\omega_{ik}\,A_{kj};
\end{equation}
where $\omega_{ij}$ is the spin:
\begin{equation}
\omega_{ij}=\frac{1}{2}\,\left({\partial v_i\over \partial x_j}-{\partial{v_j}\over \partial x_i}\right).
\end{equation}
Equation (\ref{Dtensor}) implies that we are neglecting nonlinear elasticity, and  are assuming that the elastic parts of all displacements are small.  Note, however, that we are making no such assumption about the plastic displacements.  In this Eulerian formulation, neither displacements nor strains appear explicitly in the equations of motion, and the velocity field $v_i(x,t)$ can describe arbitrarily large and complex motions of material points. 

Consistent with our assumption of constant density $\rho_0$, we assume that the plastic part of the rate-of-deformation tensor $D^{pl}_{ij}$, like $s_{ij}$, is a traceless symmetric tensor, so that plastic deformation is volume conserving. This assumption is not necessary for any theoretical purpose; it simply reduces the mathematical complexity of the analysis and is an accurate approximation in most physical circumstances.  When dilation becomes important, it is easy to add a volume nonconserving  term to $D^{pl}_{ij}$.  

\subsection{STZ Equations of Motion}
\label{STZ eom} 

The role of the STZ theory is to provide a constitutive relation between the plastic rate of deformation tensor $D_{ij}^{pl}$ and the deviatoric stress tensor $s_{ij}$.  For illustrative purposes, we focus on a common class of situations in which plastic deformation is slow on molecular time scales, and in which $D_{ij}^{pl}$ has the form
\begin{equation}
\label{D-chi-s}
\tau_0\,D_{ij}^{pl}= e^{-e_Z/\chi}\,f_{ij}({\bf s},\theta).
\end{equation}
Here, ${\bf s}$ denotes the stress tensor; $\theta = k_B\,T$ and $\chi = k_B\,T_{e\!f\!f}$ are, respectively, the ordinary and effective temperatures in energy units; $e_Z$ is an STZ formation energy; and $\tau_0$ is the molecular time scale. A low-temperature expression for $f_{ij}({\bf s},\theta)$ is shown below in Eq.(\ref{fij}). In principle, $D_{ij}^{pl}$ also depends on two internal state variables: a dimensionless STZ density $\Lambda$, and a tensor $m_{ij}$ that carries orientational memory.  Both of these variables must satisfy their own equations of motion, and both play central roles in the derivation of the STZ equations of motion in Sec.\ref{STZ}.  In going from such equations to the special form shown in Eq.(\ref{D-chi-s}), we have assumed that both $\Lambda$ and $m_{ij}$ have reached their equilibrium values in times short compared to the time scale for plastic deformation.  As will be seen, it is the dynamics of $m_{ij}$ that, at low temperatures and for quasi-static deformations, causes $f_{ij}({\bf s},\theta)$ to vanish when the magnitude of the stress $|s|$ is smaller than a dynamic yield stress $s_y$. At somewhat higher temperatures, the $m$ dynamics produces a smooth transition from thermally assisted creep to steadily driven flow at about the same yield stress.  By dropping $m_{ij}$ on the right-hand side of Eq.(\ref{D-chi-s}), we exclude consideration of Bauschinger effects or anelastic stress-strain responses below the yield stress, which are contained naturally in more general versions of this theory.

Equation (\ref{D-chi-s}) reflects the central premise of the STZ theory that, during plastic deformation, localized, irreversible molecular rearrangements take place sporadically at well separated sites.  The rearrangements that have been observed directly in numerical simulations \cite{FL98,DENNIN-OHERN-08,HAXTON-LIU07} are STZ transitions from one of their orientational states to another.  The STZ's themselves are ephemeral, noise activated, configurational fluctuations that happen to be susceptible to stress-driven, shear transformations.  An STZ, when formed, rapidly undergoes a shear transition if it is aligned favorably with respect to the stress.  Once this happens, it cannot transform further in the original direction; but it can transform backwards if the stress is reversed before significant further deformation occurs \cite{DENNIN-OHERN-08}. In either case, the STZ ultimately disappears into the background of noisy fluctuations. This picture of STZ transitions as infrequent events, and the correspondingly long time scales associated with plastic deformation, is expressed in Eq.(\ref{D-chi-s}) by small values of the Boltzmann-like factor $\exp\,(-\,e_Z/\chi)$.   

Equation (\ref{D-chi-s}) must be supplemented by an equation of motion for $\chi$.  As will be seen in Sec.\ref{eom-Teff},  this equation is fundamentally a statement about entropy flow in a driven system.  It has the form
\begin{equation}
\label{chidot0}
\tau_0\,{\dot{\chi}\over e_Z} = \kappa_1\,e^{-\,e_Z/\chi}\,\Gamma({\bf s},\theta)\,\left[1 - {\chi\over \chi_{ss}(q)}\right] + \kappa_2\, e^{-\,e_A/\chi}\,\rho(\theta)\,\left(1 - {\chi\over \theta}\right),
\end{equation} 
where $\kappa_1$ and $\kappa_2$ are dimensionless constants. The first factor on the right-hand side of Eq.(\ref{chidot0}), i.e. the product $\exp\,(-e_Z/\chi)\,\Gamma({\bf s},\theta)$, is proportional to the rate at which entropy is produced by the driving forces. We show in Sec.\ref{part2} that, at low temperatures, this product is proportional to the work rate $D_{ij}^{pl}\,s_{ij}$, which contains the factor $\exp\,(-e_Z/\chi)$ according to Eq.(\ref{D-chi-s}).  Thus, the plastic deformation and the effective temperature are similarly slow quantities.  More generally, $\Gamma({\bf s},\theta)$ is a non-negative noise strength that must be temperature dependent, because the work rate by itself can be negative if, for example, thermal fluctuations drive plastic flow in a direction opposite to the stress \cite{JSL08}.  

The second factor in the term proportional to $\kappa_1$ tells us that $\chi$ approaches a steady-state value $\chi_{ss}(q)$.  Here, $q \equiv \tau_0\,|{\bf D}^{pl}| = \tau_0\,\sqrt{(1/2)\,{\bf D}:{\bf D}}$ is the magnitude of the plastic strain rate in units of the molecular frequency $\tau_0^{-1}$.  Under normal circumstances, $q \ll 1$, and $\chi_{ss}(0) \equiv \chi_0$ is a measure of the disorder induced by slow straining or stirring. On the other hand, at very large strain rates, $q \sim 1$, and $\chi_{ss}(q)$ becomes large.  

The term proportional to $\kappa_2$ on the right-hand side of Eq.(\ref{chidot0}) is the rate at which $\chi$ relaxes to $\theta$ in the absence of external driving.  This rate contains the factor $\exp\,(-e_A/\chi)$, which determines the frequency of configurational fluctuations that couple to ordinary thermal fluctuations, in rough analogy to the way in which STZ's couple to the external stress.  In general, we expect the formation energy $e_A$ for such fluctuations to be different from $e_Z$.  The factor $\rho(\theta)$ is the attempt frequency for the thermal coupling events.  It is a super-Arrhenius function of $\theta$ that we presume to vanish below the glass transition, implying that the effective temperature is frozen into the system at sufficiently low $\theta$, where aging ceases.  

\begin{figure}
\centering \epsfig{width=.52\textwidth,file=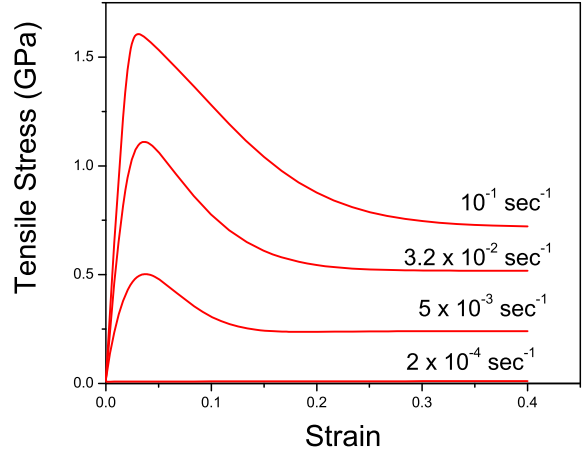} \caption{Theoretical stress-strain curves for Vitreloy 1, at 643 K, for different strain rates as shown. These curves are in good agreement with data reported by Lu et al. \cite{LUetal03}.  The one exception is that, for the top-most curve, there is no data at strains beyond the stress peak, presumably because the sample failed at that point.} \label{stress-strain}
\end{figure}

\begin{figure}
\centering \epsfig{width=.52\textwidth,file=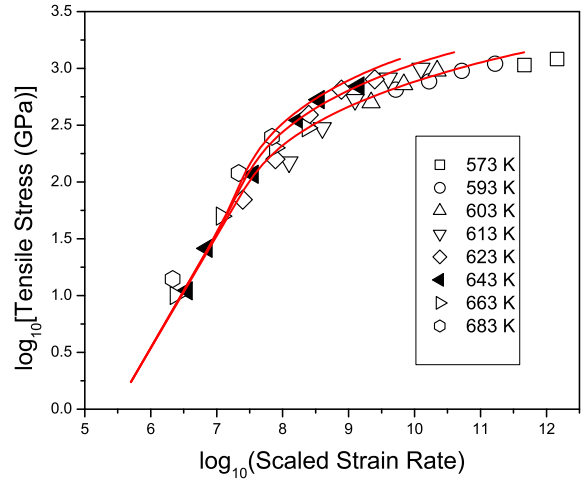} \caption{Tensile stress for Vitreloy 1 as a function of the scaled strain rate $2\,\eta_N\,\dot\gamma$, where $\eta_N$ is the Newtonian viscosity. The data points, with temperatures as indicated, are taken from Lu et al. \cite{LUetal03}. The three solid red curves, from bottom to top, are theoretical predictions for temperatures $T = 573\,K$, $643\,K$ and $683\,K$ respectively.} \label{BMGstress}
\end{figure}

\subsection{Stress-Strain Relations for a Bulk Metallic Glass}

As a first illustration of the STZ theory in operation, consider the deformation measurements carried out by Lu et al. \cite{LUetal03} using the bulk metallic glass Vitreloy 1 (${\rm Zr_{41.2}\,Ti_{13.8}\,Cu_{12.5}\,Ni_{10}\,Be_{22.5}}$).  In these experiments, a uniform bar of this material was subjected to a uniaxial compressive stress, which was measured as a function of strain over a wide range of constant strain rates, and over a range of temperatures above the glass transition.  The STZ analysis of this data is described in \cite{JSL04,JSL08}.  Here, we summarize only general features of the comparison between theory and experiment.  

Theoretical stress-strain curves for four different  homogeneous strain rates are shown in Fig.\ref{stress-strain}. This set of curves, and a similar set for different temperatures, are all in good quantitative agreement with the experimental data.  As seen here, the stress first rises elastically, proportional to the strain, while $D_{ij}^{pl}$ remains small on the right-hand side of Eq.(\ref{Dtensor}).  As $\chi$ and the density of STZ's increases according to Eq.(\ref{chidot0}), the plastic flow becomes dominant, and the stress relaxes to its steady-state value. Almost all of the STZ parameters used in plotting these curves were determined from steady-state data, including those appearing in the transition-rate formula, shown below in Eq.(\ref{R-s}), and the values of the thermal coupling factor $\rho(\theta)$ in Eq.(\ref{chidot0}), which were obtained from the measured Newtonian viscosity.  The transient behavior in Fig.\ref{stress-strain}, i.e. the crossover from elastic to plastic response, is determined primarily by the parameter $\kappa_1$ in Eq.(\ref{chidot0}) or, equivalently, the dimensionless effective specific heat $c^{e\!f\!f}$ defined below, in the text following Eq.(\ref{chidot1}).  The most important feature of these results is that, with a single fixed value of $c^{e\!f\!f}$ of the order of unity, the theory naturally reproduces the positions of the stress peaks and the rates at which these transients relax toward steady-state. In other words, the STZ theory accurately predicts the nonequilibrium dynamics of these systems, including the competition  between elastic and inelastic mechanisms, over a broad range of experimental conditions. 

In Fig.\ref{BMGstress}, we show a comparison between theory and experiment for steady-state stresses at different temperatures, as functions of the strain rate multiplied by the Newtonian viscosity.   These steady-state stresses are the same as those seen in Fig.\ref{stress-strain} in the limit of large strain. When the strain rate is scaled in this way, all of the data in the viscous limit of small stresses and strain rates automatically falls on a single curve with constant slope.  The important feature of this figure is that the curves cross over from linear viscosity to what is called ``super-plasticity'' at increasing strain rate and/or increasing temperature.  The full curves almost, but not quite, collapse onto each other; the crossover occurs at somewhat higher stress for higher temperatures in both the theoretical curves and the data. All of the temperatures shown in Fig.\ref{BMGstress} are above the glass temperature; thus the linear viscosity at small stresses can be understood as thermally assisted plastic flow.  The nonlinear response at larger stresses occurs at approximately the low-temperature yield stress, indicating that the yielding mechanism described below, following Eq.(\ref{mdot2}), becomes operative in this regime.  Thus, the quantitative agreement between theory and experiment in Fig.\ref{BMGstress} is a fairly stringent test of a central feature of the STZ theory.     

\begin{figure}
\centering \epsfig{width=.52\textwidth,file=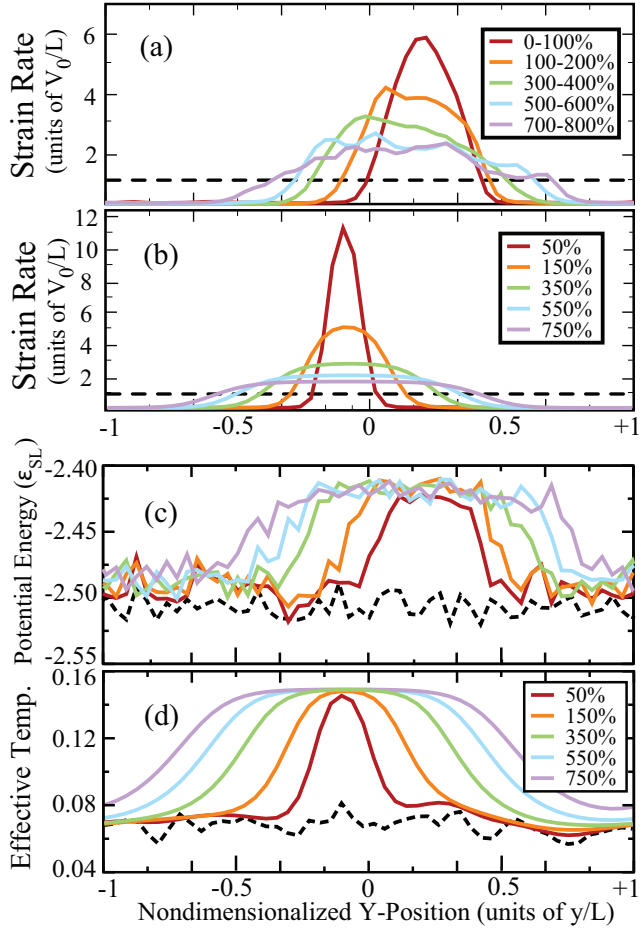} \caption{(a) Simulated strain rates,  averaged over increments of $100\%$ total strain, as functions of position at various strains.  The dashed line is the imposed average strain rate. (b) STZ predictions corresponding to the simulation data in (a).  (c) Simulated potential energy per atom as a function of position at the same total strains shown in (a).  (d) STZ predictions for the effective temperature as a function of position. The dashed lines in (c) and (d) show the initial values for the potential energy and effective temperature.} \label{bands}
\end{figure}

\subsection{Shear Banding}

One of the most important applications of STZ theory has been in explaining the mechanism of strain localization that leads to the formation of shear bands and shear fracture.  This failure mechanism is the primary reason why metallic glasses exhibit limited ductility and are not widely used as structural materials.  In steels undergoing high rates of deformation, shear localization apparently is caused by a feedback of some kind between a softening mechanism and the heat released during deformation. In our opinion, this mechanism has yet to be understood. In metallic glasses, however, localization is common even at relatively low loading rates; and it has long been suspected, and recently demonstrated convincingly \cite{LEWANDOWSKI-GREER-05}, that the instability leading to localization must be quite different from that in polycrystalline materials. The thermal conductivity of metallic glasses is too high for adiabatic heating to account for an instability on the small length scales observed experimentally.

In the STZ theory as summarized in Eq.(\ref{D-chi-s}), softening occurs due to variations in the effective temperature, which must diffuse only very slowly at rates proportional to the local shear rate.  In fact, the existence of shear bands in simulations \cite{SHI-FALK-05,SHI-FALK-06,LI-LI-05,BAILEYetal-06,CAO-CHENG-MA-09} provides an ideal virtual laboratory for testing some of the assumptions of the STZ theory.  Since the shear rate varies by orders of magnitude from inside to outside the shear band, the effective temperature also must vary significantly.  Under simple shear-loading conditions, this variation provides an opportunity to measure the deformability as a function of structure under laboratory-scale applied stresses.  Shi and Falk \cite{SHIetal07} have simulated a two-dimensional, low-temperature, binary Lennard-Jones system in order to test the Boltzmann-like relation between effective temperature and shear rate in Eq.(\ref{D-chi-s}). To do this, they used the ``quasi-thermodynamic'' assumption \cite{BLP07II} that the effective temperature is proportional to the average potential energy per atom.  Manning, et al. \cite{MANNINGetal-SHEARBANDS-07} have solved the STZ equations shown in Sec.\ref{STZ eom} above, and have found good agreement with the molecular dynamics results as well as some unexpected interpretations of them.

The comparisons between simulation \cite{SHIetal07} and theory \cite{MANNINGetal-SHEARBANDS-07} are shown in Fig.\ref{bands}.  The system is a two-dimensional strip subject to simple shear tractions imposed along the upper and lower edges.  The theory used a simplified athermal STZ transition rate $R(s)$, defined below in Eq.(\ref{Ndot}), that rises linearly at stresses appreciably larger than $s_y$.  Figures \ref{bands} (a) and (b) show, respectively, the simulated and theoretical shear rates, averaged over the length of the strip, as functions of position along the transverse direction denoted by $Y$.  As indicated, the different curves are snapshots at different total strains ranging up to $800\%$. Figs.\ref{bands} (c) and (d) show the potential energy and effective temperature as functions of position at roughly the same sequence of total strains.  In accord with the quasi-thermodynamic assumption, these sets of functions track each other accurately. 

The quantitative agreement between the simulations and theoretical results shown in Fig.\ref{bands}, along with a stability analysis in \cite{MANNINGetal-SHEARBANDS-07}, reveals that shear banding in these materials is a nonlinear, transient instability. The system is initially in a state of uniform shear indicated by the dashed horizontal lines at the bottoms of Figs.\ref{bands} (a) and (b).  The dashed curve at the bottom of Fig.\ref{bands} (c) is the initial potential energy, whose irregularity was determined by the rate at which the sample was quenched from a high temperature.  The irregular, initial effective temperature in Fig.\ref{bands} (d) was chosen to have approximately the same spatial noise spectrum as the simulated potential energy.  The band appears only when this spatial irregularity has a large enough amplitude, and then only when the strain rate and the initial average effective temperature satisfy conditions discussed in \cite{MANNINGetal-SHEARBANDS-07}.  Its position depends on the initial noise distribution; but the numerical and theoretical bands behave almost identically.  They rise rapidly and, for a while, take up almost the entire strain rate, which drops to a very small value outside the bands.  Both the potential energy and the effective temperature saturate inside the bands at limiting values corresponding to $\chi_{ss}(0) = \chi_0$ in Eq.(\ref{chidot0}).  At very late stages, when the total strain has reached multiples of $100\%$, the band slowly spreads out and collapses, because the small strain rates in the outer regions of the system slowly drive $\chi$ to its steady-state value $\chi_0$, and the entire system flows plastically.  
 
In general, the steady-state effective temperature $\chi_{ss}(q)$ rises rapidly when the dimensionless strain rate $q$ approaches unity.  According to the analysis in \cite{MANNINGetal-SHEARBANDS-09}, this property of $\chi_{ss}$ causes strongly driven shear bands to collapse, producing very narrow, fracture-like, failure zones.  Daub and coworkers \cite{DAUB-08,CARLSON-DAUB-10} have used the STZ theory to describe the dynamics of the granular material in an earthquake fault, and have shown that this fracture mechanism can account for the sudden stress drops sometimes observed in large seismic events.

\begin{figure}[here]
\centering \epsfig{width=.52\textwidth,file=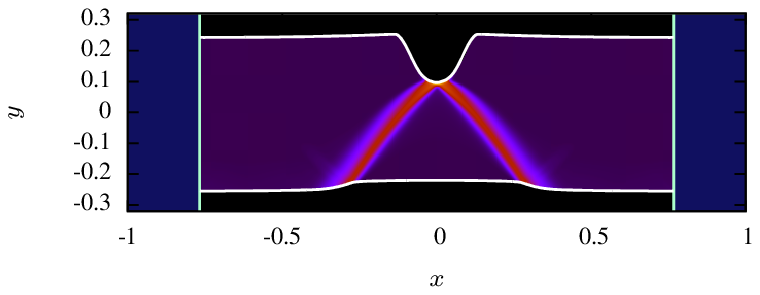}  
\end{figure}
\begin{figure}[here]
\centering \epsfig{width=.52\textwidth,file=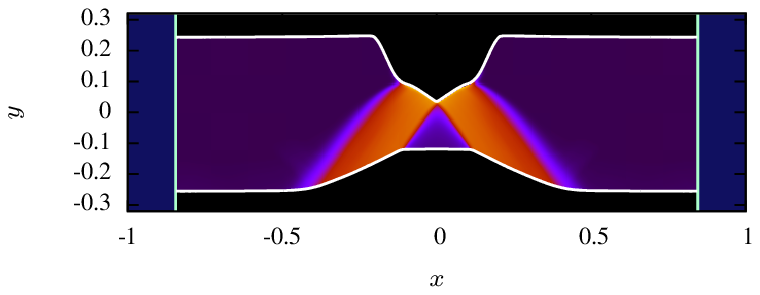} 
\end{figure}
\begin{figure}[here]
\centering \epsfig{width=.52\textwidth,file=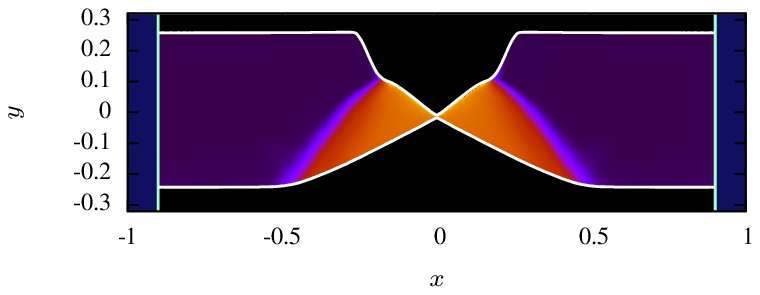}  
\end{figure}
\begin{figure}[here]
\centering \epsfig{width=.52\textwidth,file=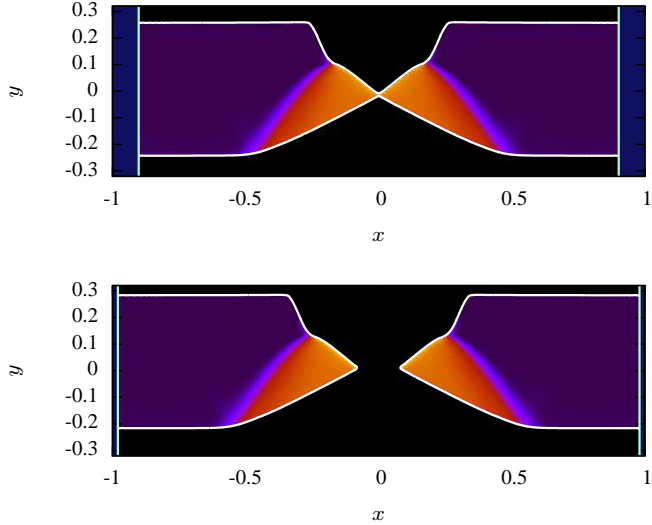} \caption{Four snapshots of a necking instability computed by Rycroft and Gibou \cite{RYCROFT-10} by solving the elasto-plastic equations of motion with STZ plasticity. The red regions are effectively hotter; i.e., they have higher effective disorder temperatures, and therefore have undergone more irreversible plastic deformation than the bluer regions.} \label{Rycroft}
\end{figure}

\subsection{Free-Boundary Problems}

Perhaps the most ambitious goal of the STZ theory is to use it in the full, elasto-plastic equations of motion shown in Sec.\ref{elastoplasticity}, and to predict time-dependent deformations of finite systems subject to external tractions.  A first, numerically unsophisticated step in this direction was made in \cite{EPL-03}. More recently, Bouchbinder and coworkers have used STZ plasticity in a studies of cavitation instabilities. \cite{CAVITY1-07,CAVITY2-08,CAVITY3-08} The computational problem is challenging, partly because including both rapid elastic and slow plastic responses in a single numerical procedure is difficult, and especially because this is necessarily a free-boundary problem in which the geometry is changing as a function of time.  

Figure \ref{Rycroft} shows recent results by Rycroft and Gibou \cite{RYCROFT-10}, in which a necking instability is followed all the way to fracture.  The system is a two-dimensional strip subject to  tractions exerted by inflexible, vertically sliding grips on the left- and right-hand sides, which move away from each other at a fixed speed.  The strip initially has a smooth notch near the center of its upper edge.  The red and blue regions indicate higher and lower effective temperatures respectively.  In the top picture, a pair of effectively hotter, i.e. internally disordered, shear bands emerges from the notch along the directions of maximum shear stress.  These bands, like the one shown in Fig.\ref{bands}, then broaden into slipping regions. Ultimately, the strip separates into two parts, each with a pattern of residual disorder in the places where the local plastic flow was largest.  Other results, not shown here, indicate that elastic energy initially is stored uniformly throughout the strip. Then, as the necking instability grows, this energy flows to the neck and is dissipated there. 

A detailed description of the numerical procedure used to generate these pictures can be found in \cite{RYCROFT-10}.  Very briefly -- The simulation was based on an athermal STZ theory of the kind described here in Sec.\ref{part2}, with a very simple rate factor $R(s)$ similar to that used in the shear-banding analysis described above. The equations of motion were the two-dimensional versions of those shown in Sec.\ref{elastoplasticity}, except that a small viscous term proportional to $\nabla^2 v_i$ was added to the right-hand side of Eq.(\ref{acceleration}) in order to damp out elastic oscillations.   The boundary was tracked using a level-set method. The upper and lower edges were free surfaces; there was no surface tension.  

The main limitation of this numerical scheme is that, so far, it has been useful only for describing ductile behavior of the kind seen in these pictures.  Had the model been brittle, or perhaps had it been numerically possible to explore substantially larger pulling speeds with the same model parameters, one or more cracks might have started at the notch and propagated downward through the system.  If this technical limitation can be overcome, we should have a powerful tool for studying dynamic fracture. 

\section{Thermodynamic Derivation of the STZ Equations}
\label{part2}

\subsection{Athermal STZ Model}

For simplicity, in this part of the review, we focus primarily on what we call the ``athermal'' limit of plasticity theory.  By ``athermal,'' we do not mean strictly zero temperature.  On the contrary, as stated in Sec.\ref{overview}, we assume that there is always some thermal or mechanical noise that sets the time scale $\tau_0$ for rapid, small-scale motions.  However, we assume that this noise is not strong enough to cause large-scale molecular rearrangements in the absence of external forcing.  In particular, $\rho(\theta) = 0$ in Eq.(\ref{chidot0}).  As a result, models of this kind have well defined yield stresses, but not linear viscosities, and they do not exhibit thermally induced strain recovery.  They do describe, for example, irreversible deformation of glasses below or near the glass temperature, or the flow of densely packed granular materials subject to stresses large enough that they become unjammed. At the end of this Section, we discuss briefly how the athermal theory has been supplemented to produce the more general equations of motion shown in Sec.\ref{part1}. 

Our thermodynamic analysis is based on the assumption that the configurational degrees of freedom within a solidlike material are driven out of equilibrium with the heat bath when the system is persistently deformed by external forcing, and that they are naturally described by an effective temperature under those circumstances.\cite{BLII-09}  We use the term ``configurational degrees of freedom'' to denote the mechanically stable molecular positions that change slowly during irreversible deformation, as opposed to the much faster molecular vibrations about the stable configurations. Mathematically, the configurational degrees of freedom specify the ``inherent structures''  \cite{GOLDSTEIN-69,STILLINGER-WEBER-82,STILLINGER-88}.  Formation of STZ's, or STZ transitions between their internal orientational states, are events in which the system moves from one inherent structure to another. 

\subsection{First and Second Laws of Thermodynamics for a Plastic Solid}
\label{thermo}

The preceding discussion implies that an amorphous, solidlike material consists of two, weakly coupled subsystems: the slow configurational degrees of freedom on the one hand, and the fast kinetic-vibrational degrees of freedom on the other. The fast degrees of freedom are strongly coupled to a heat bath so that, together, they and the heat bath constitute a thermal reservoir at temperature $\theta = k_B\,T$.  

It is useful to start with a microcanonical formulation in which the energy $U_C$ of the configurational subsystem is a function of its entropy $S_C$, its volume $V$, an elastic shear strain ${\bf \epsilon}$, and a set of internal variables $\{\Lambda\}$ that in Sec.\ref{STZ} will become the number density of STZ's and a measure of their average orientation.  Throughout this discussion, we choose entropies to be dimensionless quantities -- logarithms of numbers of states -- and express the temperatures $\theta$ and $\chi$ in units of energy.  For simplicity, consider only pure shear deformation in, say, the $x,y$ plane, so that the deviatoric stress tensor has components $s_{xx} = -\,s_{yy} = s$, the elastic strain tensor is $\epsilon_{xx} = -\,\epsilon_{yy} = \epsilon$, and the rate of plastic deformation tensor is $D^{pl}_{xx} = -\,D^{pl}_{yy} = D^{pl}$.  Let the thermal reservoir have energy $U_R$ and entropy $S_R$.  This reservoir has none of its own internal degrees of freedom, and does not support a shear stress. According to the definition of temperature, the effective temperature of the configurational subsystem is
\begin{equation}
\chi = \left({\partial U_C\over \partial S_C}\right)_{\epsilon,\{\Lambda\}},
\end{equation}
which is not necessarily the same as $\theta = \partial U_R/\partial S_R$. 

The total energy of this system is
\begin{equation}
U^{tot} = U_C(S_C,{\bf \epsilon},\{\Lambda\}) + U_R(S_R).
\end{equation}
The first law of thermodynamics,
\begin{equation}
\label{firstlaw1}
2\,V\,s\,D^{tot} = \dot U^{tot},
\end{equation}
says simply that energy is conserved when work is done on the system at the rate $2\,V\,s\,D^{tot}$. Assume, as in Eq.(\ref{Dtensor}), that the total rate of deformation $D^{tot}$ is the sum of elastic and plastic parts, i.e. $D^{tot} = \dot \epsilon + D^{pl}$.  If $V\,s = (\partial U_C/\partial \epsilon)_{S_C,\{\Lambda\}}$, i.e. if the stress is wholly elastic in origin, then the elastic terms cancel out on either side of Eq.(\ref{firstlaw1}).  We therefore omit $\epsilon$, as well as the constant volume $V$, as explicit arguments of $U_C$ and $S_C$.  The first law becomes
\begin{equation}
\label{firstlaw2}
2\,V\,s\,D^{pl} = \chi\,\dot S_C + \sum_{\alpha}\left({\partial U_C\over \partial \Lambda_{\alpha}}\right)_{S_C} \dot\Lambda_{\alpha} + \theta\,\dot S_R.
\end{equation}

The fundamental statistical statement of the second law of thermodynamics is that the total entropy of an isolated system is a non-decreasing function of time:
\begin{equation}
\label{secondlaw1}
\dot S_{tot} = \dot S_C + \dot S_R \ge 0.
\end{equation}
As argued in \cite{BLI-09}, this statement is thermodynamically self-consistent only if the set $\{\Lambda\}$ consists of a small number of state variables, each of which is an extensive quantity (or the volume average of such a quantity).  Using Eq.(\ref{firstlaw2}) to evaluate $\chi\,\dot S_C$ in Eq.(\ref{secondlaw1}), we find
\begin{equation}
\label{secondlaw2}
{\cal W}(s,\{\Lambda\})+ (\chi - \theta)\,\dot S_R \ge 0; ~~~~{\cal W}(s,\{\Lambda\})\equiv 2\,V\,s\,D^{pl} - \sum_{\alpha}\left({\partial U_C\over \partial \Lambda_{\alpha}}\right)_{S_C} \dot\Lambda_{\alpha}.
\end{equation}
This inequality must be satisfied for arbitrary, independent variations of the $\Lambda_{\alpha}$ and $S_R$; thus each of its component terms must separately be non-negative.  We immediately enforce $(\chi - \theta)\,\dot S_R \ge 0$ by writing
\begin{equation}
\dot U_R = \theta\,\dot S_R = A(\chi,\theta)\,(\chi - \theta)\equiv -\,Q,
\end{equation}
where $A(\chi,\theta)$ is a non-negative thermal conductivity, and $Q$ is the rate at which heat is flowing from the thermal reservoir into the configurational degrees of freedom.  

The inequality ${\cal W}(s,\{\Lambda\})\ge 0$ is a form of a Clausius-Duhem inequality that requires a non-negative rate of heat production; that is, the rate at which work is done must exceed the rate at which energy is stored internally.  We will use this inequality in Sec.\ref{STZ} to deduce features of the STZ equations of motion. Note that we have derived this inequality from fundamental principles, using an unambiguous statistical definition of the entropy \cite{BLI-09}, rather than postulating it as an axiomatic form of the second law.  The latter strategy is the one that is common in the literature.  See, for example, the monographs by Lubliner \cite{LUBLINER-90}, Maugin \cite{MAUGIN-99}, and Nemat-Nasser \cite{NEMAT-NASSER-04}; or the classic series of studies by by Coleman, Noll, and Gurtin  \cite{COLEMAN-NOLL-63,COLEMAN-GURTIN-67}.

\subsection{Equation of Motion for the Effective Temperature}
\label{eom-Teff}

Our first-law equation, Eq.(\ref{firstlaw2}), now has the form
\begin{equation}
\label{firstlaw3}
\chi\,\dot S_C = {\cal W}(s,\{\Lambda\}) + Q.
\end{equation}
We can use Eq.(\ref{firstlaw3}) to derive an equation of motion for $\chi$ by making several  observations.  First, although the STZ's account for all of the coupling beween the applied stress and the plastic deformation, they are very rare fluctuations and constitute only a negligibly small fraction of the total energy or entropy of the configurational subsystem.  Thus Eq.(\ref{firstlaw3}) is a simple statement of energy conservation that can be reduced to
\begin{equation}
\label{chidot1}
V\,c^{e\!f\!f}\,\dot\chi \approx 2\,V\,s\,D^{pl} +Q.
\end{equation}
where $V\,c^{e\!f\!f} = \chi\,(\partial S_C/\partial \chi)$ is the effective heat capacity.  

Second, note that the only relevant rate factor in this athermal system is the work rate $2\,V\,s\,D^{pl}$ itself.  So long as there are no thermal fluctuations capable of inducing reverse plastic flow, $D^{pl}$ must have the same sign as $s$, and this rate is non-negative.  Moreover, in the absence of such fluctuations, $Q$ must be proportional to $2\,V\,s\,D^{pl}$; the configurational system does not move at all without external forcing.  

Third, we know that $\chi$ must reach some steady-state value during steady shear flow.  As in Eq.(\ref{chidot0}), define the dimensionless strain rate $q \equiv \tau_0\,|D^{pl}|$, and denote the steady-state effective temperature by $\chi_{ss}(q)$.  (See \cite{JSL-MANNING-TEFF-07} for a detailed discussion of the $q$ dependence of $\chi_{ss}$.) In the limit $q\ll 1$ for an athermal amorphous system, $\chi_0 = \chi_{ss}(0)$ is roughly (perhaps exactly) equal to the glass transition temperature, i.e. $\chi_0 \sim k_B\,T_g$.  In other words, athermal systems reach fluctuating steady states of disorder under slow shear.  The slower the shear, the longer the system takes in real time to reach steady state; but the ultimate value of $\chi_{ss}$ must be independent of $q$ simply for dimensional reasons -- there are supposedly no competing time scales when $q \to 0$.  By definition, the right-hand side of Eq.({\ref{chidot1}) vanishes when $\chi = \chi_{ss}(q)$.  Therefore, for $\chi$ not too far from $\chi_{ss}(q)$, we approximate Eq.(\ref{chidot1}) by
\begin{equation}
\label{chidot2}
c^{e\!f\!f}\,\dot\chi \approx 2\,s\,D^{pl}\,\left[1 - {\chi\over\chi_{ss}(q)}\right].
\end{equation}
Here we see explicitly that the characteristic time scale for $\chi$ is the same as the time scale for plastic deformation, and both are slow because $D^{pl}$ is proportional to the small density of STZ's.  

\subsection{STZ equations of motion}
\label{STZ}

We turn now to constructing an athermal STZ model based on effective-temperature thermodynamics.

It is easiest and physically most transparent to assume that the STZ's are oriented only in the $\pm$ directions relative to the stress.  In fact, we lose no generality by doing this; the tensorial generalizations of the equations are obvious at the end of the analysis.  Let the number of $\pm$ STZ's be $N_{\pm}$, and let the total number of molecular sites be $N$.  The master equation for the $N_{\pm}$ is
\begin{equation}
\label{Ndot}
\tau_0\,\dot N_{\pm} = R(\pm s)\,N_{\mp} - R(\mp s)\,N_{\pm} + \Gamma(s)\,\left({N_{eq}\over 2} - N_{\pm}\right),
\end{equation}
where $R(\pm s)/\tau_0$ is the rate factor for STZ transitions between their orientations, and $\Gamma(s)/\tau_0$ is the corresponding factor for noise driven creation and annihilation of STZ's.    The equilibrium number $N_{eq}$ and the rate factor $\Gamma(s)$ will be determined shortly by thermodynamic arguments.  The internal state variables $\Lambda_{\alpha}$ introduced in Sec.\ref{thermo} are
\begin{equation}
\label{Lambdadef}
\Lambda = {N_++N_-\over N};~~~~m ={N_+-N_-\over N_++N_-}.
\end{equation} 
Here, $\Lambda$ is the fractional density of STZ's; and $m$ is their orientational bias which, as mentioned following Eq.(\ref{D-chi-s}), becomes the traceless, symmetric tensor $m_{ij}$ in more general versions of the theory.  According to Eq.(\ref{Ndot}), the equations of motion for $\Lambda$ and $m$ are
\begin{equation}
\label{Lambdadot}
\tau_0\,\dot \Lambda = \Gamma\,(\Lambda_{eq} - \Lambda);~~~~ \Lambda_{eq} = {N_{eq}\over N};
\end{equation}
and
\begin{equation}
\label{mdot}
\tau_0\,\dot m = 2\,{\cal C}(s)\Bigl[{\cal T}(s) - m\Bigr] - \Gamma\,m -{\tau_0\,\dot\Lambda\over \Lambda}\,m,
\end{equation}
where
\begin{equation}
{\cal C}(s)= {1\over 2}\,\Bigl[R(s) + R(-s)\Bigr];~~~~{\cal T}(s)= {R(s) - R(-s)\over R(s) + R(-s)}.
\end{equation}
The rate of plastic deformation is 
\begin{equation}
\tau_0\,D^{pl} = {v_0\over V}\,\Bigl[R(s)\,N_--R(-s)\,N_+\Bigr] = \epsilon_0\,\Lambda\,{\cal C}(s)\Bigl[{\cal T}(s) - m\Bigr],
\end{equation}
where $v_0$ is a molecular-scale volume that sets the size of the plastic strain increment induced by an STZ transition. We expect  $\epsilon_0 \equiv N\,v_0/V$ to be a number of the order of unity.

Our model of rare, noninteracting STZ's implies that we can write the entropy in the form
\begin{equation}
\label{SC}
S_C(U_C,\Lambda,m) = N\,S_0(\Lambda) + N\,\Lambda\,\psi(m) + S_1(U_1),
\end{equation}
where $S_1$ and $U_1$ are, respectively, the entropy and energy of all the non-STZ degrees of freedom; $\psi(m)$ is the internal entropy associated with STZ's of average orientation $m$; and, for $\Lambda \ll 1$,
\begin{equation}
S_0(\Lambda) \cong - \,\Lambda\,\ln\,\Lambda + \Lambda.
\end{equation}
With this assumption, the configurational energy becomes
\begin{eqnarray}
\label{UC}
U_C(S_C,\Lambda,m) &=& N\,\Lambda\,e_Z + U_1(S_1)\cr  &=& N\,\Lambda\,e_Z + U_1\Bigl(S_C - NS_0(\Lambda) - N\,\Lambda\,\psi(m)\Bigr).
\end{eqnarray}
We now evaluate the partial derivatives of $U_C$ in Eq.(\ref{secondlaw2}), obtaining
\begin{eqnarray}
\label{secondlaw3}
\nonumber
{\tau_0\over N}\,{\cal W}(s,\Lambda,m) &=& -\,{\partial F_Z\over \partial \Lambda}\,\tau_0\,\dot\Lambda - \Gamma\,\chi\,\Lambda\,m\,{\partial\psi\over\partial m}\cr \\ &+&  2\,\Lambda\,{\cal C}(s)\Bigl[{\cal T}(s) - m\Bigr]\left[s\,v_0+ \chi\,{\partial \psi\over\partial m}\right] \ge 0,
\end{eqnarray}
where
\begin{equation}
F_Z(\Lambda,m) = e_Z\,\Lambda - \chi\,S_0(\Lambda) - \chi\,\Lambda\,\left[\psi(m) - m\,{\partial \psi\over\partial m}\right].
\end{equation}

As before, the three terms in this inequality must separately be non-negative; but the argument, especially for the third term, is nontrivial.   The term proportional to $\dot\Lambda$ is non-negative if
\begin{equation}
\tau_0\,\dot\Lambda \propto -\,{\partial F_Z\over \partial \Lambda},
\end{equation}
or, more generally, if $\Lambda$ has a dynamical fixed point at a minimum of the free-energy-like function $F_Z$.  This minimum occurs at 
\begin{equation}
\label{Lambdaeq}
\Lambda = \Lambda_{eq} = \nu(m)\, e^{-\,e_Z/\chi};~~~~~\nu(m) = \exp\,\left[\psi(m)-m\,{\partial \psi\over\partial m}\right],
\end{equation}
which is consistent with the definition of  $\Lambda_{eq}$ in Eq.(\ref{Lambdadot}). 

The internal entropy $\psi(m)$ is necessarily a positive, symmetric function with a maximum at $m=0$; therefore, the second term in Eq.(\ref{secondlaw3}) is automatically non-negative given a properly chosen $\psi(m)$. 

The last term in Eq.(\ref{secondlaw3}) is the most interesting because, unlike the $\dot\Lambda$ term, this inequality does not lead to a free-energy minimization law.  Nor does it imply normal flow in a free-energy landscape as advocated in \cite{LUBLINER-90} or \cite{NEMAT-NASSER-04}.  It is the only one of the three terms in Eq.(\ref{secondlaw3}) that depends explicitly on the stress $s$, which can, in principle, be assigned any value independent of $\Lambda$ or $m$.  This term can be made to be non-negative for all values of $s$, and for $-1 < m < 1$, by choosing
\begin{equation}
\label{xidef}
{\partial \psi\over\partial m}= -\,{v_0\over\chi}\,\xi(m),
\end{equation}
where $\xi(m)$ is the functional inverse of ${\cal T}(s)$; that is, ${\cal T}\Bigl(\xi(m)\Bigr) = m$. This choice means that both $s$-dependent factors in this product are monotonic functions that vanish at the same $m$-dependent value of $s$.  We can use this second-law constraint in either of two ways.  In \cite{BLIII-09}, it was assumed that the STZ's were strictly two-state systems with no internal degrees of freedom, and therefore had an Ising-like entropy.  In that case, the rate factor $R(s)$ had to be proportional to $\exp\,(v_0\,s/\chi)$.  A more realistic interpretation is that the STZ's are complex, many-body systems with many internal degrees of freedom.  The better strategy, then, is to choose a physically motivated form of $R(s)$ and to let that determine $\psi(m)$ via the choice of Eq.(\ref{xidef}).  

The latter strategy works especially well in the athermal limit that we are considering here.  The physically most important feature of that limit is that the rearrangement transitions always go in the direction of the stress; the noise is not strong enough to drive them in the opposite direction.  This means that $R(-|s|) \ll R(+|s|)$, and ${\cal T}(s) \approx {\rm sign}\,(s)$.
Equation (\ref{xidef}) then implies that, for $-1<m<1$, $\partial \psi/\partial m \approx 0$ and, in Eq.(\ref{Lambdaeq}),
\begin{equation}
\label{nudef}
\nu(m) \approx \nu(0) = \exp\,[\psi(0)]
\end{equation}
is the number of molecules in an STZ. Interestingly, the athermal choice of $R(s)$, via the second law of thermodynamics, recognizes that ergodicity is broken on time scales relevant to STZ transitions.  Equation (\ref{nudef}) implies that any given molecular site has $\nu(0)$ different ways of being part of an STZ of size $\nu(0)$, independent of the average STZ orientation $m$.  This can be true only if the STZ is not switching back and forth between its orientations during the time over which we are averaging to compute $m$.  

Now return to Eq.(\ref{secondlaw3}).  With our athermal assumption, only the last term in the expression for ${\cal W}$ remains nonzero.  The term proportional to $\dot\Lambda$ vanishes for  slow deformations; and the second term vanishes because $\partial \psi/\partial m \approx 0$.  Up to a factor with the dimensions of energy, the quantity ${\cal W}$ is the non-negative rate at which configurational entropy is being generated.  It was first argued by Pechenik \cite{LP03,PECHENIK05}  that the noise strength $\Gamma$ should be proportional to this rate of entropy generation per STZ, with the proportionality factor, say $v_0\,s_0$, necessarily having the dimensions of energy.  Therefore,
\begin{equation}
\label{Gammadef}
\Gamma\,\Lambda\,v_0\,s_0 = {\tau_0\over N}\,{\cal W} \approx 2\,\Lambda\,{\cal C}(s)\Bigl[{\cal T}(s) - m\Bigr]v_0\,s;~~~~\Gamma \approx 2\,{\cal C}(s)\Bigl[{\cal T}(s) - m\Bigr]\,{s\over s_0}.
\end{equation}
The resulting relation between the STZ production rate, $\Gamma(s)\,N_{eq}/\tau_0$ in Eq.(\ref{Ndot}), and the work rate $2\,s\,D^{pl}$ was guessed in \cite{FL98}, and has been confirmed by Heggen {\it et al.} \cite{HEGGENetal05} in the context of conventional flow-defect theories. This identification of the rate at which configurational disorder is created with the strength of mechanically generated noise has proved to be a very useful concept, as will be seen below in evaluating the yield stress in Eq.(\ref{Dathermal}).  

The equation of motion for $m$, Eq.(\ref{mdot}), with $\dot\Lambda \approx 0$, becomes
\begin{equation}
\label{mdot2}
\tau_0\,\dot m \approx  2\,{\cal C}(s)\Bigl[{\cal T}(s) - m\Bigr]\,\left( 1 - {s\,m\over s_0}\right).
\end{equation}
Note that both Eqs.(\ref{Lambdadot}) for $\dot\Lambda$ and (\ref{mdot2}) for $\dot m$ describe relaxation to steady state that is fast compared to that of the effective temperature described by Eq.(\ref{chidot2}).  The factor $D^{pl}$ on the right-hand side of Eq.(\ref{chidot2}) contains the small factor $\Lambda$; but no such factor appears in Eq.(\ref{Lambdadot}) or (\ref{mdot2}).  Thus, we confirm that the STZ variables $\Lambda$ and $m$ are dynamically slaved to relatively slow changes in $s$ and $\chi$, which is the assumption that we used in deriving Eqs.(\ref{D-chi-s}) and (\ref{chidot0}). 

Equation (\ref{mdot2})  is the usual STZ theory result. There is an exchange of stability at the stress $s = s_0$. For $|s| < s_0$, the dynamically stable, steady-state solution of Eq.(\ref{mdot2}) is the jammed state with $m \approx \pm 1$; and the rate of deformation $D^{pl}$ is zero. On the other hand, for $|s|> s_0$, the stable solution is $m = s_0/s$, and 
\begin{equation}
\label{Dathermal}
\tau_0\,D^{pl} \approx \epsilon_0\,\nu(0)\,e^{-\,e_Z/\chi}\,{\cal C}(s)\,\left[{\rm sign}(s) - {s_0\over s}\right]
\end{equation}
Thus $s_0= s_y$ is the dynamic yield stress.  

To complete the derivation, we need to choose the rate factor $R(s)$.  One possibility that has worked well in several applications is a thermally activated rate of the form:
\begin{equation}
\label{R-s}
R(s) = R_0(s)\,\exp\,\left[-\,{\Delta(s)\over \theta}\right];~~~~\Delta(s) = \Delta_0\,e^{-\,s/\bar\mu}.
\end{equation}
where $R_0(s)$ is a symmetric function of the stress.  The exponential form of the barrier height $\Delta(s)$ is the simplest possible expression that vanishes for large positive stress, diverges at large negative stress, and introduces only a single new parameter $\bar\mu$. For  the metallic glass calculations in \cite{JSL08}, $R_0(s) = \sqrt{1+(s/s_1)^2}$ with $s_1 \sim s_y$.  Equation (\ref{R-s}) is consistent with the athermal approximation if $\theta \ll\Delta_0$.  

Two generalizations of the athermal equations derived above are needed in order to recover the fully thermal STZ theory shown in Eqs.(\ref{D-chi-s}) and (\ref{chidot0}).  So long as we are dealing with an isotropic material, the only directional information in the system is contained in the deviatoric stress.  We can then assume that the plastic rate-of-deformation tensor is proportional to $s_{ij}/|s|$, where $|s| = \sqrt{(1/2)\,s_{ij}s_{ij}}$.  Thus, using Eq.(\ref{Dathermal}), we find
\begin{equation}
\label{fij}
f_{ij}({\bf s}) \approx \epsilon_0\,\nu(0)\,{s_{ij}\over |s|}\,{\cal C}(|s|)\,\left(1 - {s_0\over |s|}\right).
\end{equation}
Equation (\ref{chidot2}), the athermal version of Eq.(\ref{chidot0}), becomes
\begin{equation}
\label{chidot3}
c^{e\!f\!f}\,\dot\chi \approx s_{ij}\,D^{pl}_{ij}\,\left[1 - {\chi\over\chi_{ss}(q)}\right].
\end{equation}
with $q = \tau_0\,|D^{pl}|$. Further generalizing these results to fully thermal situations is straightforward but considerably more complicated.  The essential step is to recognize that the mechanical noise strength $\Gamma$ introduced in Eq.(\ref{Ndot}) must become the sum of incoherent mechanical plus thermal noise strengths, i.e. $\Gamma \to \Gamma + \rho(\theta)$, where $\rho(\theta)$ is the same thermal term that we introduced in Eq.(\ref{chidot0}) to account for relaxation in the absence of mechanical driving forces.  The explicit form of $\Gamma$ can be computed using the same second-law argument that led to Eq.(\ref{Gammadef}).  Details can be found in \cite{JSL08}.

\section{Concluding Remarks}
\label{conclusions}

So far as we know, the STZ theory is the only existing mathematical description of solidlike amorphous plasticity that starts with realistic molecular models, and uses the principles of nonequilibrium thermodynamics to guide the prediction of observed phenomena.  To date, those phenomena have included: the transition between linear viscosity and superplasticity as a function of temperature and strain rate for bulk metallic glasses; the transient as well as steady-state parts of the stress-strain curves for real and numerically simulated glass-forming materials; the nature of transient shear-banding instabilities in glassy materials; and even a quantitative understanding of the granular shearing instability that produces sharp stress drops during major earthquakes.  

The formulation of nonequilibrium thermodynamics that emerged during the development of the STZ theory \cite{BLI-09,BLII-09,BLIII-09} recently has been extended to a study of memory effects in thermally cycled glass formers, i.e. the Kovacs effect \cite{BL-Kovacs-10}.  This thermodynamic point of view even has provided an accurate account of a remarkably wide range of experimental data for dislocation-mediated plasticity in polycrystalline solids \cite{LBL-10}.  The similarities and differences between dislocations and STZ's are interesting in themselves.  Dislocations are well defined entities, directly observable by electron microscopy and subject to fairly well understood, deterministic equations of motion.  It is only when large numbers of interacting dislocations are driven by external forces into chaotic motion that thermodynamic concepts become relevant to them.  

In contrast, the STZ's have never enjoyed the visibility of dislocations.  The elementary rearrangements presumably associated with flow defects have been known for decades; but, for systems in the process of deformation, it  never has been possible to identify the defects themselves before the events occurred. The thermodynamic theory developed here implies that, with perhaps a few special exceptions, such prior identifications are impossible for most practical purposes.  In the present theory, the sequence of STZ creation, shear transition, and annihilation is a noise activated process, more nearly akin to nucleation of a critical droplet in a supercooled vapor than, for example, to the creation of a dislocation at a Frank-Read source.  We should no more expect to be able to look at a deforming amorphous material and predict where the next STZ event will occur than we should expect to be able to predict where the next droplet will form in the vapor.  Nor should we worry that the stochastic nature of STZ plasticity unnecessarily limits the predictive power of the theory.  

The one important case where a deterministic, dynamical theory of amorphous deformation should make sense is in the athermal quasistatic (AQS) limit.  At zero temperature, using numerical simulation, we might be able to strain an amorphous system gradually, and predict where the next rearrangement will occur by looking at nearby saddle points in the energy landscape.  Once the system has crossed a saddle point, however, we cannot predict where the next such event will occur unless we stop straining the system and let it relax into its nearest energy minimum before resuming the deformation.  This is the AQS numerical procedure, which often produces system-spanning, avalanche-like events and size-dependent noise spectra.  There are many real systems that do behave like this, for example, granular materials, foams, or colloidal suspensions sheared so slowly that the mechanical noise generated by one event has died out before the next event occurs.  These are not what we would call ``normal'' plastic materials; they cannot be described by local constitutive laws like the ones discussed here.  But it may  be interesting to locate the boundary between normal and AQS systems, and thus to understand the limitations of the STZ theory. 

In our opinion, however, it will be more interesting to use the theoretical tools developed here to explore ``normal'' plasticity in broader contexts -- in particular, to study a variety of dense, complex fluids and biological materials. We need to understand the relations between the STZ and SGR theories, and perhaps learn how to combine the strengths of the two approaches. We have some new tools for understanding nonequilibrium phenomena; and we are optimistic that these tools will lead us to new discoveries.  

\begin{acknowledgments}
We thank Eran Bouchbinder and Michael Cates for reading early versions of this paper and for making many valuable suggestions.  We also thank C. Rycroft and F. Gibou for providing the pictures shown in Fig.\ref{Rycroft} prior to their publication.  M.L. Falk acknowledges support from the U.S. National Science Foundation under Award DMR0808704. J.S. Langer acknowledges support from U.S. Department of Energy Grant No. DE-FG03-99ER45762. 
\end{acknowledgments}

\end{document}